\documentclass[prl,aps,twocolumn]{revtex4-1}

\usepackage[pdftex]{graphicx}
\usepackage{amsbsy,amssymb,amsmath,bm,mathtools}

\usepackage{xspace}
\usepackage{hyperref}

\usepackage{bm}
\usepackage{color}

\begin{document}


\title{Plaquette order in classical spin liquid stabilized by strong off-diagonal exchange}

\author{Preetha Saha}

\author{Zhijie Fan}

\author{Depei Zhang}

\author{Gia-Wei Chern}
\affiliation{Department of Physics, University of Virginia, Charlottesville, VA 22904, USA}

\date{\today}

\begin{abstract}
We report a new classical spin liquid in which the collective flux degrees of freedom break the translation symmetry of the honeycomb lattice. This exotic phase exists in frustrated spin-orbit magnets where a dominant off-diagonal exchange, the so-called $\Gamma$ term, results in a macroscopic ground-state degeneracy at the classical level. We demonstrate that the system undergoes a phase transition driven by thermal order-by-disorder at a critical temperature $T_c \approx 0.04 |\Gamma|$. At first sight, this transition reduces an emergent spherical spin-symmetry to a cubic one: spins point predominantly toward the cubic axes at $T < T_c$. However, this seems to simply restore the cubic symmetry of the $\Gamma$ model, and the non-coplanar spins remain disordered below~$T_c$. We show that the phase transition actually corresponds to plaquette ordering of hexagonal fluxes and the cubic symmetry is indeed broken, a scenario that is further confirmed by our extensive Monte Carlo simulations. 
\end{abstract}

\maketitle

Mott insulators with strong spin-orbit coupling have generated considerable interest recently~\cite{krempa14}. The local magnetic degrees of freedom in such materials are entities with significant orbital character. This special property leads to effective interactions that exhibit strong anisotropy in both real and pseudo-spin spaces, as described by novel Hamiltonians such as quantum compass or 120$^\circ$ models~\cite{kugel82,nussinov15}.  A new type of magnetic frustration~\cite{wu08,zhao08,chern11}, which is different from the well studied geometrical frustration~\cite{chalker92,moessner98,lacroix11}, originates from the nontrivial interplay between lattice geometry and anisotropic spin-orbital exchange. One recent representative example is the spin-1/2 honeycomb Kitaev model~\cite{kitaev06} with Ising-like interactions involving different spin components on the three distinct nearest-neighbor bonds. Remarkably, the Kitaev model is exactly solvable and exhibits a quantum spin-liquid ground state with fractionalized excitations~\cite{kitaev06,knolle14,nasu14}. The classical limit of the Kitaev Hamiltonian also exhibits a macroscopic ground-state degeneracy and interesting order-by-disorder phenomena~\cite{baskaran08,chandra10,price12,rousochatzakis17b}. 

The recent enormous interest in frustrated spin-orbit magnets is triggered by the realization that spin interactions in certain $4d$ and $5d$ Mott insulators are dominated by the anisotropic Kitaev-type exchange~\cite{jackeli09,chaloupka10,chaloupka13,trebst17}. 
The presence of other spin interactions, notably the isotropic Heisenberg exchange, in these compounds eventually drives the system into a magnetically ordered state despite a dominate Kitaev term~\cite{singh12,choi12,ye12,chun15,williams16,plumb14,sears15,kubota15,johnson15,biffin14,biffin14b,modic14}. Nevertheless, the search for spin liquids in frustrated spin-orbit magnets continues. Experimentally, tuning spin interactions by applying magnetic field~\cite{zheng17,wang17,baek17} or pressure~\cite{takayama15,veiga17} has been attempted to suppress the magnetic order. On the theoretical side, it has been pointed out that the off-diagonal exchange anisotropy, the so-called $\Gamma$ term, plays a crucial role in the magnetic behaviors of these spin-orbit Mott insulators~\cite{sizyuk14,rau14,winter16,kim16,nishimoto16}. In fact, the suppression of long-range order in some compounds is suspected to be due to the increased strength of $\Gamma$ interaction, instead of the enhanced Kitaev-type exchange~\cite{kim16b,majumder18}. This experimental tendency can be understood from a recent theoretical work that shows a new classical spin liquid in the idealized $\Gamma$ model on the honeycomb lattice and its three-dimensional variants~\cite{rousochatzakis17}. 

In this paper, we investigate the thermodynamic behavior of the $\Gamma$ model at low temperatures. With the aid of extensive Monte Carlo (MC) simulations, we demonstrate a phase transition at $T_c \approx 0.04\,|\Gamma|$ driven by thermal order-by-disorder. Below the transition temperature, spins mainly point toward the cubic directions, yet remain disordered. The nature of the transition at $T_c$ seems puzzling as the phase transition simply restores the cubic symmetry of the Hamiltonian by breaking an emergent spherical symmetry in the high-$T$ phase. We resolve this issue by identifying the appropriate order parameter and show that the phase at $T < T_c$ is characterized by a hidden order, in which the hexagonal fluxes spontaneously break both translation and rotation symmetries of the lattice. The cubic spin-symmetry is also broken as a result of spin-orbit coupling. Finally, we employ the Landau-Lifshitz dynamics method to study the dynamical response of spins at low temperatures. Our results show that the two spin-liquid phases, above and below $T_c$, exhibit rather distinct dynamical structure factors.

\begin{figure}
\includegraphics[width=0.99\columnwidth]{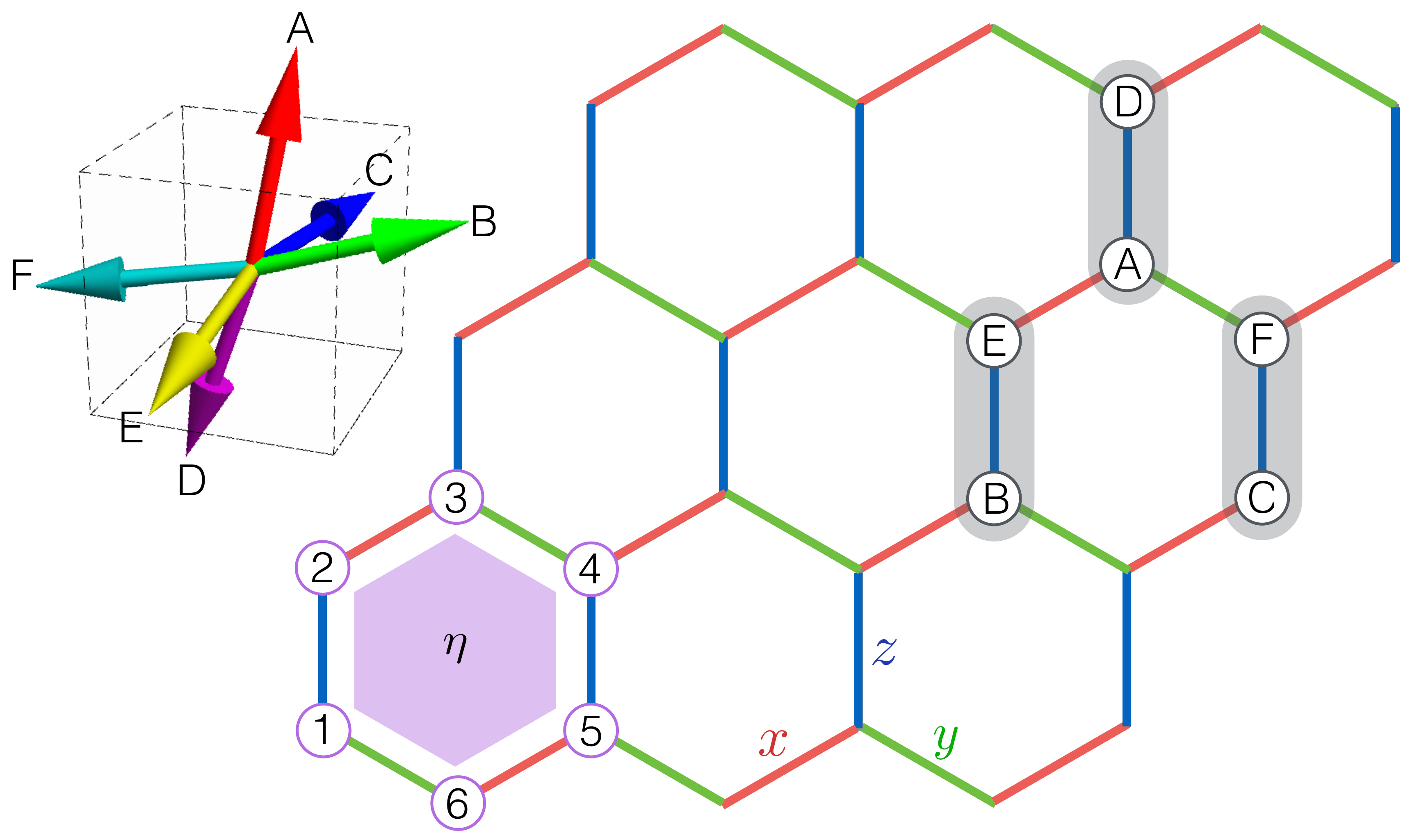}
\caption{(Color online)  
\label{fig:honeycomb} Ground states of the $\Gamma$ model on a honeycomb lattice. A generic ground state is characterized by a directional vector $\hat {\mathbf n} = (a, b, c)$ and a set of Ising variables $\{ \eta_\alpha \}$ defined on individual hexagons. To construct a ground state, first we build a perfect $\sqrt{3}\times\!\sqrt{3}$ order based on the six inequivalent spins of the tripled unit cell: $\mathbf S_{\rm A} = (a, b, c)$, $\mathbf S_{\rm B} = (c, a, b)$, $\mathbf S_{\rm C} = (b, c, a)$, $\mathbf S_{\rm D} = \zeta (b, a, c)$, $\mathbf S_{\rm E} = \zeta (a, c, b)$, and $\mathbf S_{\rm F} = \zeta (c, b, a)$. Here $\zeta = -{\rm sgn}(\Gamma)$. Next, go through every hexagon and modify the component of its six spins: $S^x_1 \to \eta S^x_1$, $S^y_2 \to \eta S^y_2$, $S^z_3 \to \eta S^z_3$, $S^x_4 \to \eta S^x_4$, $S^y_5 \to \eta S^y_5$, and $S^z_6 \to \eta S^z_6$. In the example shown above, $\zeta = -1$.
}
\end{figure}

We consider the $\Gamma$ model on the honeycomb lattice, in which nearest-neighbor (NN) spin interaction is dominated by the off-diagonal exchange term. It involves different spin-components on the three inequivalent NN bonds, denoted as $x$, $y$, and $z$ (see Fig.~\ref{fig:honeycomb}), on the honeycomb lattice:
\begin{eqnarray}
	\label{eq:H_Gamma}
	\mathcal{H} &=& \Gamma \sum_{\langle ij \rangle \parallel x} (S^y_i S^z_j + S^z_i S^y_j) + \Gamma \sum_{\langle ij \rangle \parallel y} (S^z_i S^x_j + S^x_i S^z_j) \nonumber \\
	& & \quad + \Gamma \sum_{\langle ij \rangle \parallel z} (S^x_i S^y_j + S^y_i S^x_j).
\end{eqnarray}
Here $\langle ij \rangle \parallel \xi$ denotes NN pairs along bond of type-$\xi$. Both signs of $\Gamma$ are considered here, although energetically the two cases are equivalent due to the bipartite nature of honeycomb lattice.

\begin{figure}[t]
\includegraphics[width=0.99\columnwidth]{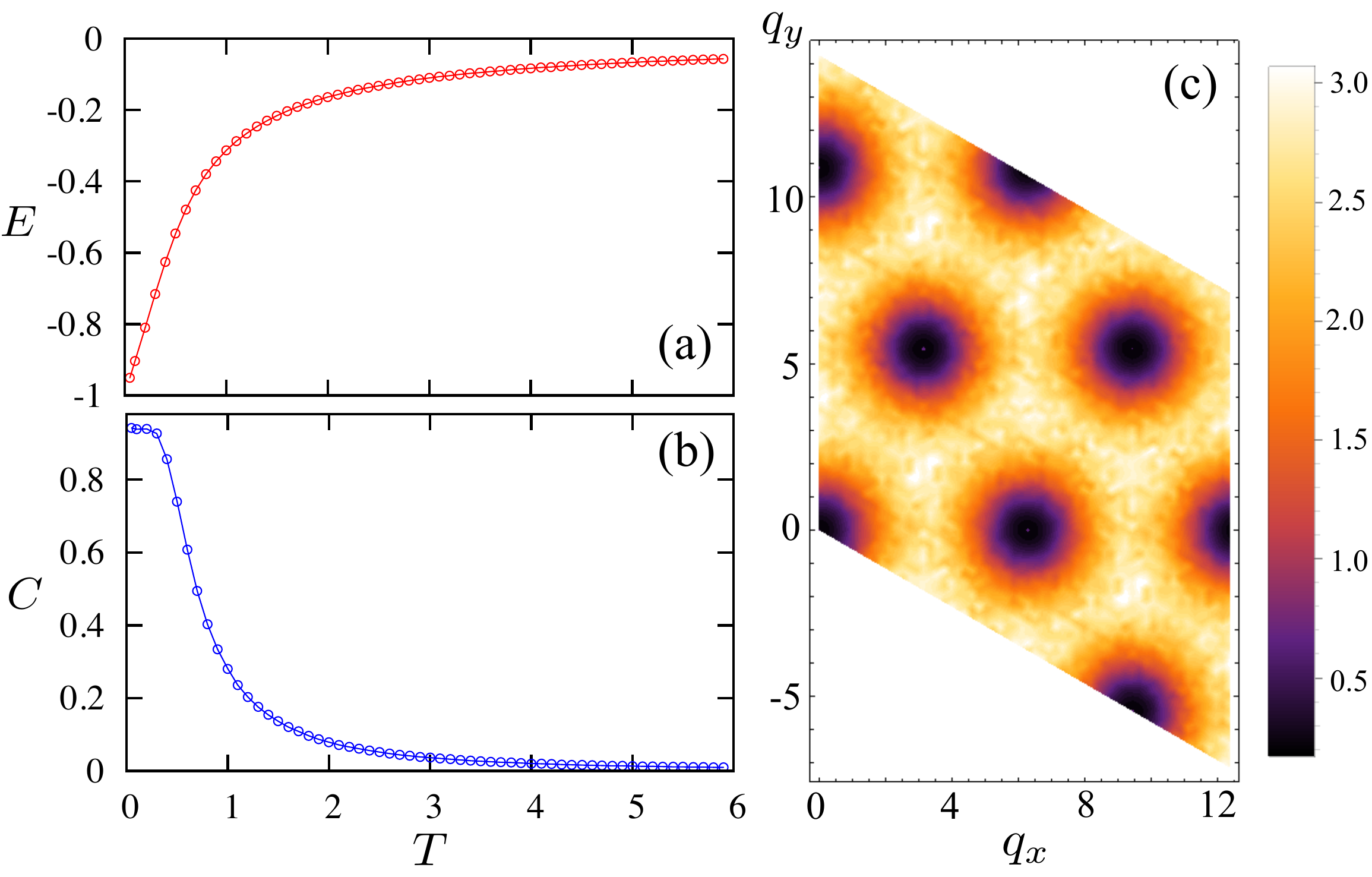}
\caption{(Color online)  
\label{fig:mc1} (a) Energy density and (b) specific heat vs temperature of the $\Gamma$ model down to temperature $T \sim 0.05$. Here both energy and temperatures are expressed in units of $|\Gamma|$. (c) shows the static structure factor of the antiferromagnetic $\Gamma$ model at $T = 0.05$. 
}
\end{figure}

The classical ground states of $\Gamma$ model are extensively degenerate~\cite{rousochatzakis17}, giving rise to a new type of classical spin liquid which is different from the familiar cases in geometrically frustrated magnets. Our MC simulations over a wide temperature range, summarized in Fig.~\ref{fig:mc1}, show no sign of phase transition down to $T \sim 0.05 |\Gamma|$. The energy density gradually approaches its minimum $E_0 = -|\Gamma|$, while the specific heat shows a plateau-like feature at $T \lesssim 0.1\,|\Gamma|$. The static structure factor of the $\Gamma > 0$ case at $T = 0.05 |\Gamma|$ exhibits a broad minima at $\mathbf q = 0$; see Fig.~\ref{fig:mc1}(c). The absence of Bragg peaks is consistent with the picture of a classical spin liquid. The fact that there is no pinch-point singularity, which is a unique feature of spin liquids in geometrically frustrated magnets~\cite{garanin99,isakov04,henley05}, also points to a different nature of the macroscopic degeneracy in $\Gamma$ model. MC simulations further find extremely short-ranged spin-spin correlation, which is similar to that seen in Kitaev spin liquid~\cite{kitaev06,baskaran08}, but different from that of geometrically frustrated systems. 

The characterization of the degenerate ground-state manifold has been discussed in great detail in Ref.~\cite{rousochatzakis17}. A generic ground state is specified by a directional vector $\hat{\mathbf n} = (a, b, c)$ and a set of Ising variables $\{\eta_\alpha \}$ defined on individual hexagons; see Fig.~\ref{fig:honeycomb}. In the classical limit, or without consideration of dynamical effects, these Ising variables are pure gauge degrees of freedom and will remain disordered at all temperatures. 
An explicit procedure for constructing the ground state is as follows. First, we use the unit vector $\hat{\mathbf n}$ to derive six inequivalent spins $\mathbf S_{\rm A}$, $\mathbf S_{\rm B}$, $\cdots$, $\mathbf S_{\rm F}$ for the tripled unit cell of a perfect $\sqrt{3} \times \!\sqrt{3}$ long-range order. Next, we go through every hexagon $\alpha$ in this periodic structure and modify the spin components: $S^x_1 \to \eta S^x_1$, $S^y_2 \to \eta S^y_2$, $S^z_3 \to \eta S^z_3$, $S^x_4 \to \eta S^x_4$, $S^y_5 \to \eta S^y_5$, and $S^z_6 \to \eta S^z_6$, where $\mathbf S_{1, \cdots, 6}$ are the six spins surrounding the $\alpha$-th hexagon. Note that the eight directions $(\pm a, \pm b, \pm c)$ correspond to the same $\hat{\mathbf n}$ as they are related by flipping the $\eta$ variable. It is thus similar to the director in nematic liquid crystal.

Since different ground states are labeled by discrete Ising variables $\{\eta_\alpha\}$, it raises the question whether the ground-state manifold is fully connected. The issue here is how one can move from one ground state {\em continuously} to another, as simply changing $\eta$ requires flipping spin component which is a discrete process. It turns out continuous transformation of $\{\eta_\alpha\}$ can be achieved with the aid of the directional vector $\hat{\mathbf n} = (a, b, c)$. To see this, we first note that each $\eta_\alpha$ is associated with only {\em one} component of the unit vector $\hat{\mathbf n}$ in the ground state. Take the hexagon shown in Fig.~\ref{fig:honeycomb} as an example, the six spins are $\mathbf S_1= (a,b,c)$, $\mathbf S_2 = \zeta (b, a, c)$, $\mathbf S_3 = (b, c, a)$, $\mathbf S_4 = \zeta(a, c, b)$, $\mathbf S_5 = (c, a, b)$, and $\mathbf S_6 = \zeta (c, b, a)$. According to the ground-state rule, the local $\eta$ only controls the `$a$'-component of the six spins in this hexagon.
As a result, all $\eta$-variables can be divided into three groups: type-A (respectively, B and C) for spin-components controlled by $a$ (respectively, $b$ and $c$). When one of the component of $\hat{\mathbf n}$ vanishes, 1/3 of the $\eta$ becomes idle. 

This feature allows us to construct a continuous path from one set of $\eta$ to another one $\eta'$. Specifically, we rotate the directional vector according to the sequence: $(a,b,c) \to (0, b', c') \to (a'', 0, c'') \to (a''', b''', 0) \to (a, b, c)$. After the first rotation, the vanishing $a$ component allows us to change 1/3 of the $\eta$ variables (those associated with $a$-component) to their counterpart in $\eta'$. Repeating similar process for the other two sets of $\eta$ then completes the transformation from $\eta$ to $\eta'$ while keeping the $\hat{\mathbf n}$ vector in the same direction.

\begin{figure}[t]
\includegraphics[width=0.92\columnwidth]{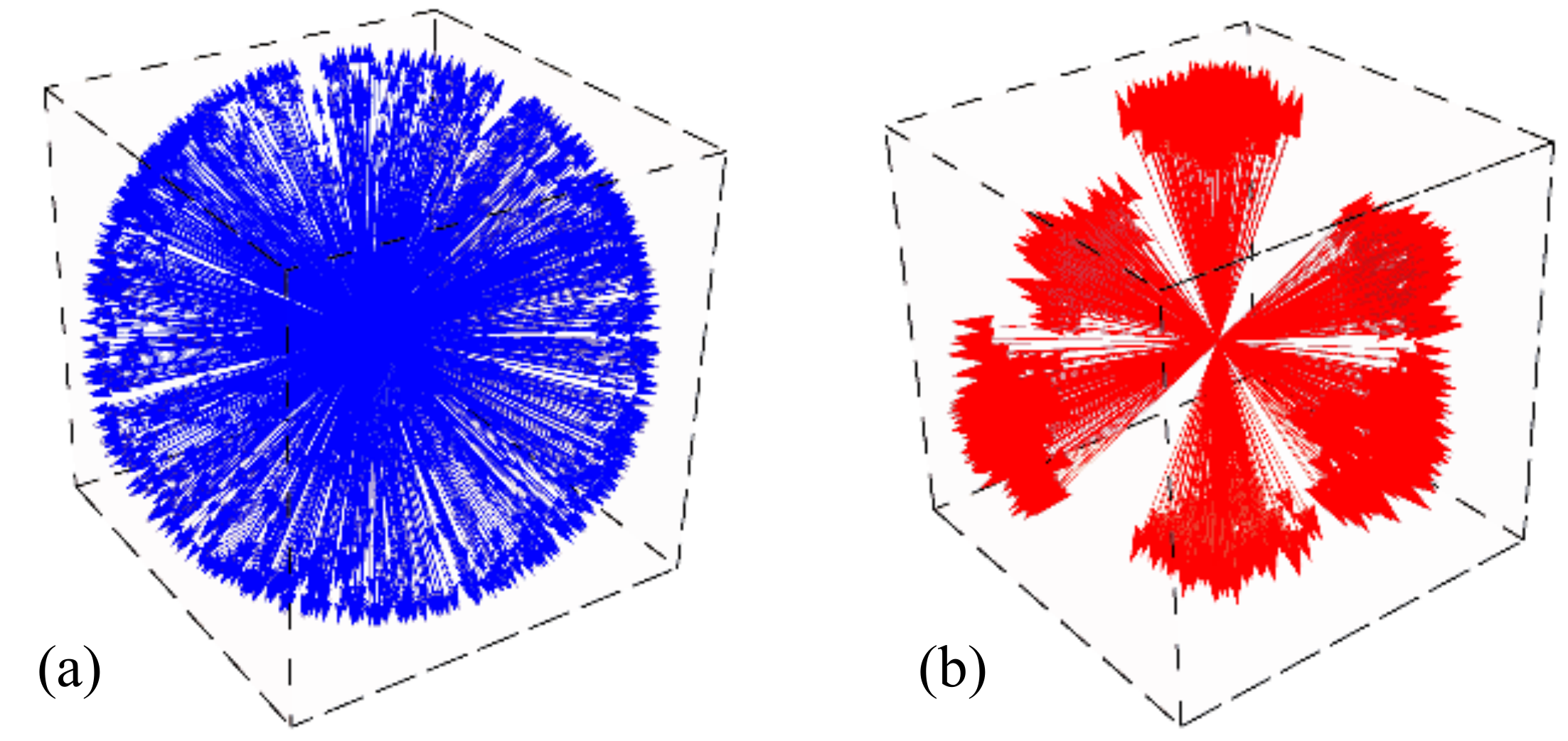}
\caption{(Color online)  
\label{fig:snapshots} Snapshots of spin configurations above and below $T_c = 0.0401 |\Gamma|$: (a) $T = 0.05$ and (b) $T = 0.03$. In the low-$T$ phase, spins predominately point toward the six cubic directions. 
}
\end{figure}

The above discussion clearly shows that the rotational symmetry of the vector $\hat{\mathbf n}$ is crucial to the connectivity of the ground-state manifold. Without this feature, different $\{\eta_\alpha\}$ becomes disjoint from each other. Interestingly, our MC simulations find a freezing phenomenon of the vector $\hat{\mathbf n}$ at a very low temperature $T_c \approx 0.04\,|\Gamma|$. This is illustrated by snapshots of spins above and below this critical temperature; see Fig.~\ref{fig:snapshots}. At $T > T_c$, the spins and the directional vector $\hat{\mathbf n}$ exhibit an emergent spherical symmetry even at temperatures well below the exchange energy scale $|\Gamma|$.  This rotational symmetry is lost at the critical temperature, and spins mainly point toward the six cubic axes, or equivalently the directional vector freezes to one of the cubic directions, i.e. $\hat{\mathbf n} \sim (1, 0, 0)$, $(0, 1, 0)$, or $(0, 0, 1)$ at $T < T_c$. As states parameterized by different $\hat{\mathbf n}$ are degenerate at the mean-field level, the cubic directions are selected by thermal fluctuations through the order-by-disorder mechanism. Equivalently, this can also be viewed as the entropic selection, resulting from an effective free energy $\mathcal{F}_{\rm ani} \propto - (a^4 + b^4 + c^4)$. Indeed, simple analysis shows that these cubic directions allow for the largest number of zero modes at the harmonic level. We note that similar cubic anisotropy is also generated by quantum fluctuations~\cite{rousochatzakis17}.

It is crucial to note that although the spin-symmetry is seemingly reduced from spherical to cubic when crossing $T_c$, this cannot be viewed as a true reduction of symmetries as the $\Gamma$ model itself is already cubic-symmetric. The apparent spherical symmetry at $T_c < T < |\Gamma|$ is an {\em emergent} property of the phase, which is due to the spatial fluctuations of directional vector $\hat{\mathbf n}(\mathbf r)$.  Another important observation is that while the degeneracy associated with vector $\hat{\mathbf n}$ is lifted by thermal fluctuation, a discrete macroscopic degeneracy persists due to the Ising gauge symmetry of $\{\eta_\alpha\}$, especially for classical spins. Consequently, spins remain disordered at $T < T_c$.

\begin{figure}[t]
\includegraphics[width=0.99\columnwidth]{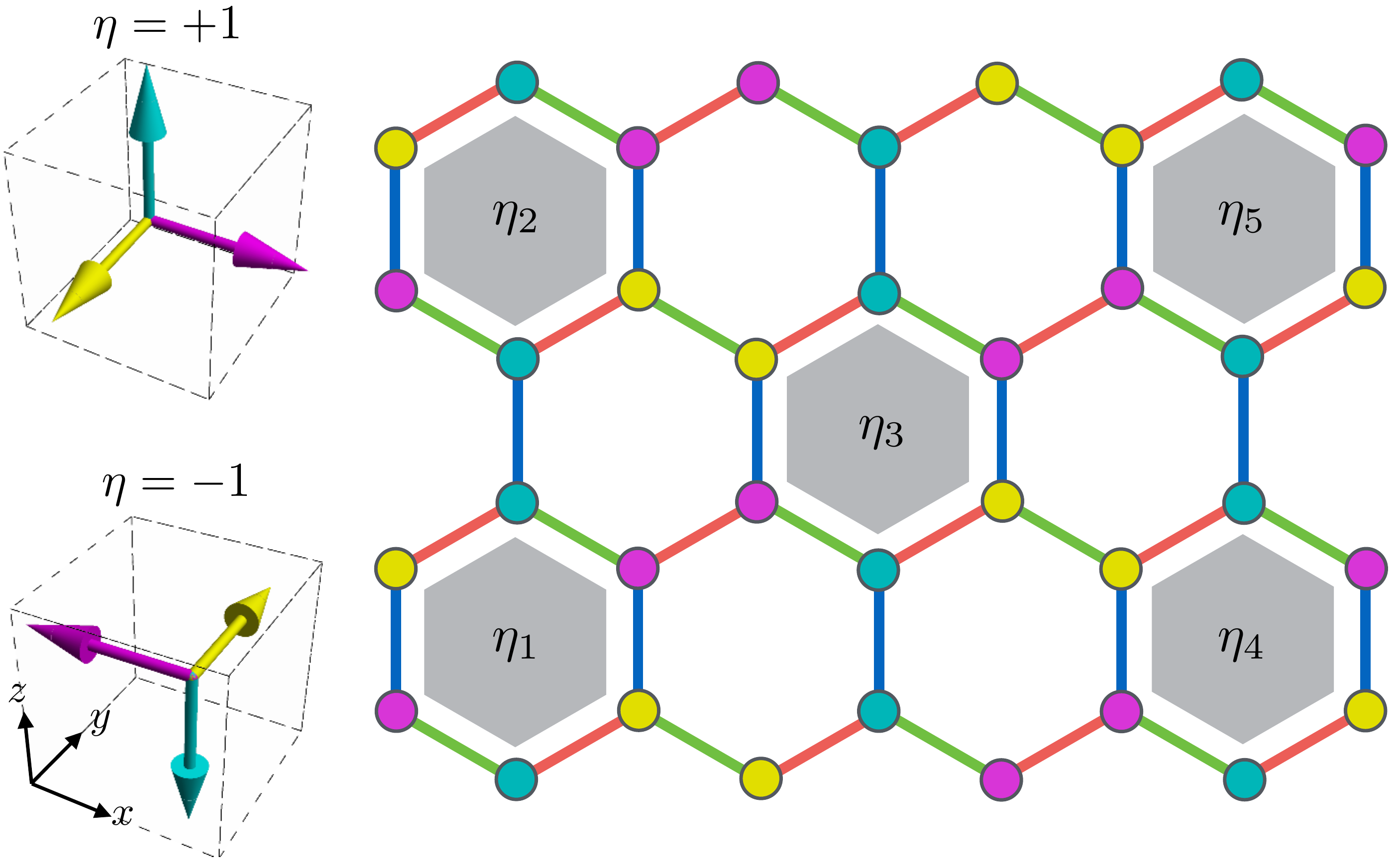}
\caption{(Color online)  
\label{fig:cubic} Plaquette order of hexagonal fluxes on honeycomb lattice. Shaded hexagons have nonzero flux $W \sim 1$, while empty hexagons have a vanishing $W$. Spins are orthogonal to each other (with left handedness for $\zeta = -1$) on each shaded hexagon; their specific directions depend on the local $\eta$, as specified in the insets. The arrangement of hexagons with finite $W$ corresponds to the famous $\sqrt{3} \times\!\sqrt{3}$ long-range order. Spins remain disordered due to uncorrelated $\eta_\alpha$ on the shaded hexagons.
}
\end{figure}

To resolve this issue and investigate the nature of the low-$T$ phase, we first note that the cubic spin-orbital symmetry of the $\Gamma$ model is indeed broken below $T_c$, yet in a complicated pattern: local spins have to pick one of the six cubic directions in a coordinated way while preserving the gauge symmetry of $\{\eta_\alpha\}$. A convenient local quantity to characterize the broken symmetry is the flux variable defined on each hexagon~\cite{kitaev06}:
\begin{eqnarray}
	W_{\alpha} = S^x_1 \,S^y_2 \, S^z_3 \, S^x_4 \, S^y_5 \, S^z_6, 
\end{eqnarray}
where $\mathbf S_{1, \cdots, 6}$ are the six spins around the $\alpha$ hexagon.  These fluxes play an important role in the spin-1/2 Kitaev model as they are ``integrals of motion'' of the Kitaev Hamiltonian~\cite{kitaev06}. In our case, the flux $W_\alpha$ is similarly a gauge-invariant variable, that is independent of $\eta_\alpha$. On the other hand, it can be used to characterize the ordering of $\hat{\mathbf n}$. To see this, we note that in the ground state, they only take on three different values~\cite{rousochatzakis17}: $W_{\rm A} = \zeta a^6$ for hexagons whose $\eta$ is associated with component $a$, and similarly $W_{\rm B} = \zeta b^6$ and $W_{\rm C} = \zeta c^6$ for the other two sets of hexagons, where $\zeta = -{\rm sgn}(\Gamma)$
As the vector $\hat{\mathbf n}$ freezes to one of the cubic directions, 2/3 of the fluxes also vanish. Since hexagons of a given type form an enlarged triangular lattice, the flux patten of the low-$T$ phase, e.g. $W_A \approx 1$, and $W_B \approx W_C \approx 0$, corresponds to a broken translation symmetry; see Fig.~\ref{fig:cubic}. Importantly, the uncorrelated $\eta_\alpha$ on hexagons with nonzero $W$ give rise to a disordered spin configuration. 
We note in passing that plaquette orders with similar spatial pattern also exist as ground state in $J_1$-$J_2$ quantum $S=1/2$ and $S=1$ honeycomb Heisenberg model~\cite{ganesh13,zhu13,zhao12,gong15}. Our finding shows a rare example of plaqutte ordering hidden in a classical spin liquid on honeycomb lattice.

\begin{figure}[t]
\includegraphics[width=0.9\columnwidth]{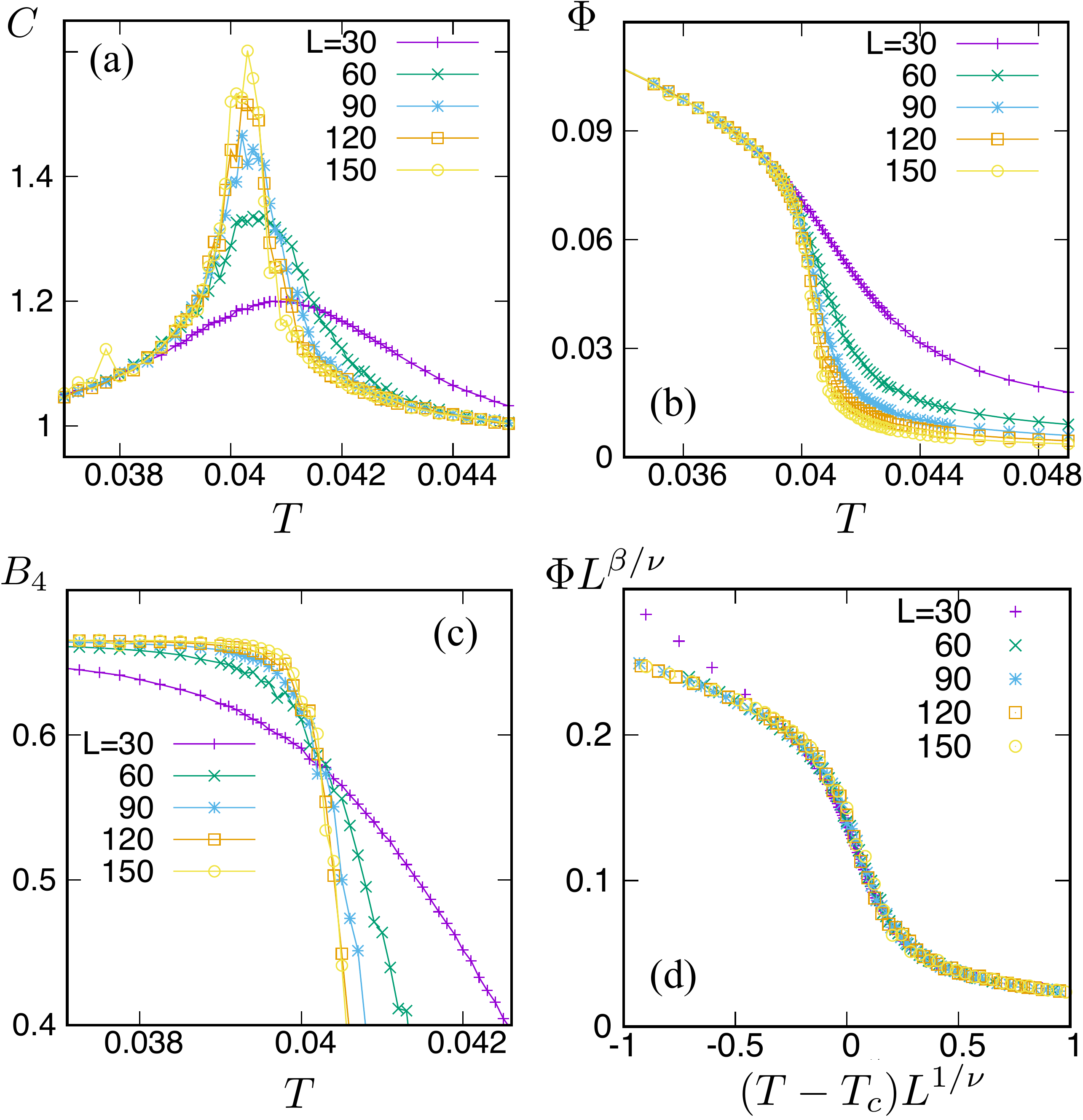}
\caption{(Color online)  
\label{fig:mc2} Monte Carlo simulation of the translation symmetry breaking of flux variables. (a) specific heat, (b) order parameter $\Phi = \langle |\tilde W(\mathbf Q)| \rangle$, (c) the corresponding Binder cumulant $B_4 \equiv 1 - \langle|\tilde W(\mathbf Q)|^4\rangle/ 3 \langle |\tilde W(\mathbf Q)|^2 \rangle^2$ as functions of temperature.  The crossing point of the Binder curves gives an estimate of $T_c \approx 0.0402$. Critical exponents of the transition are obtained from finite-size scaling: $\alpha = 0.167$, $\beta = 0.177$, $\gamma = 1.47$, and $\nu = 0.863$. For example, panel~(d) shows the data collapsing of scaled order parameter and temperature.
}
\end{figure}

The $\sqrt{3}\times\!\sqrt{3}$ arrangement of hexagons with nonzero~$W$ shown in Fig.~\ref{fig:cubic} suggests an order parameter
\begin{eqnarray}
	\tilde W(\mathbf Q) = \frac{1}{N} \sum_\alpha W_\alpha \,e^{i \mathbf Q \cdot \mathbf r_\alpha},
\end{eqnarray}
which is the Fourier transform at wavevector $\mathbf Q = (4\pi / 3, 0)$, for characterizing the broken translation symmetry. 
We then performed extensive large-scale Monte Carlo simulations at temperatures around the critical $T_c$; the results are summarized in Fig.~\ref{fig:mc2}. The specific heat shows clear finite-size effect, as the peak in $C$ grows with increasing lattice size. Moreover, the order parameter defined as $\Phi \equiv \langle |\tilde W(\mathbf Q)| \rangle$ exhibits characteristics of a second-order phase transition. For example, the growth of $\Phi$ below $T_c$ becomes sharper as $L$ is increased. More evidence of a continuous phase transition is provided by the temperature dependence of Binder cumulant $B_4 \equiv 1 - \langle|\tilde W(\mathbf Q)|^4\rangle/ 3 \langle |\tilde W(\mathbf Q)|^2 \rangle^2$ shown in Fig.~\ref{fig:mc2}(c). The various $B_4$ curves from different lattice sizes cross at a critical point $T_c \approx 0.0402 |\Gamma|$.

Finally, we also conducted detailed finite-size scaling analysis on various thermodynamics quantities to extract the critical exponents of the phase transition. For example, fair data points collapsing is obtained for the order-parameter curves; see Fig.~\ref{fig:mc2}(d). Our results demonstrate a clear second-order transition caused by thermal order-by-disorder, in stark contrast to the famous coplanar order-by-disorder in kagome antiferromagnet~\cite{chalker92,chern13}. Since coplanar spins are characterized by a vector order parameter, which cannot be spontaneously broken in 2D, the coplanar spin-ordering in kagome ends up in a crossover. On the other hand, the flux-ordering in our case corresponds to a broken $Z_3$ symmetry, i.e. which cubic direction is selected for the vector $\hat{\mathbf n}$. Some of the critical exponents, e.g. $\nu$ and $\gamma$, obtained from finite-size scaling, shown in caption of Fig.~\ref{fig:mc2}, are consistent with the 2D 3-state Potts universality class, although others show noticeable deviations. This discrepancy could be due to the gauge degrees of freedom~$\{\eta_\alpha\}$, which might have nontrivial effects on the critical behavior.

\begin{figure}[t]
\includegraphics[width=0.99\columnwidth]{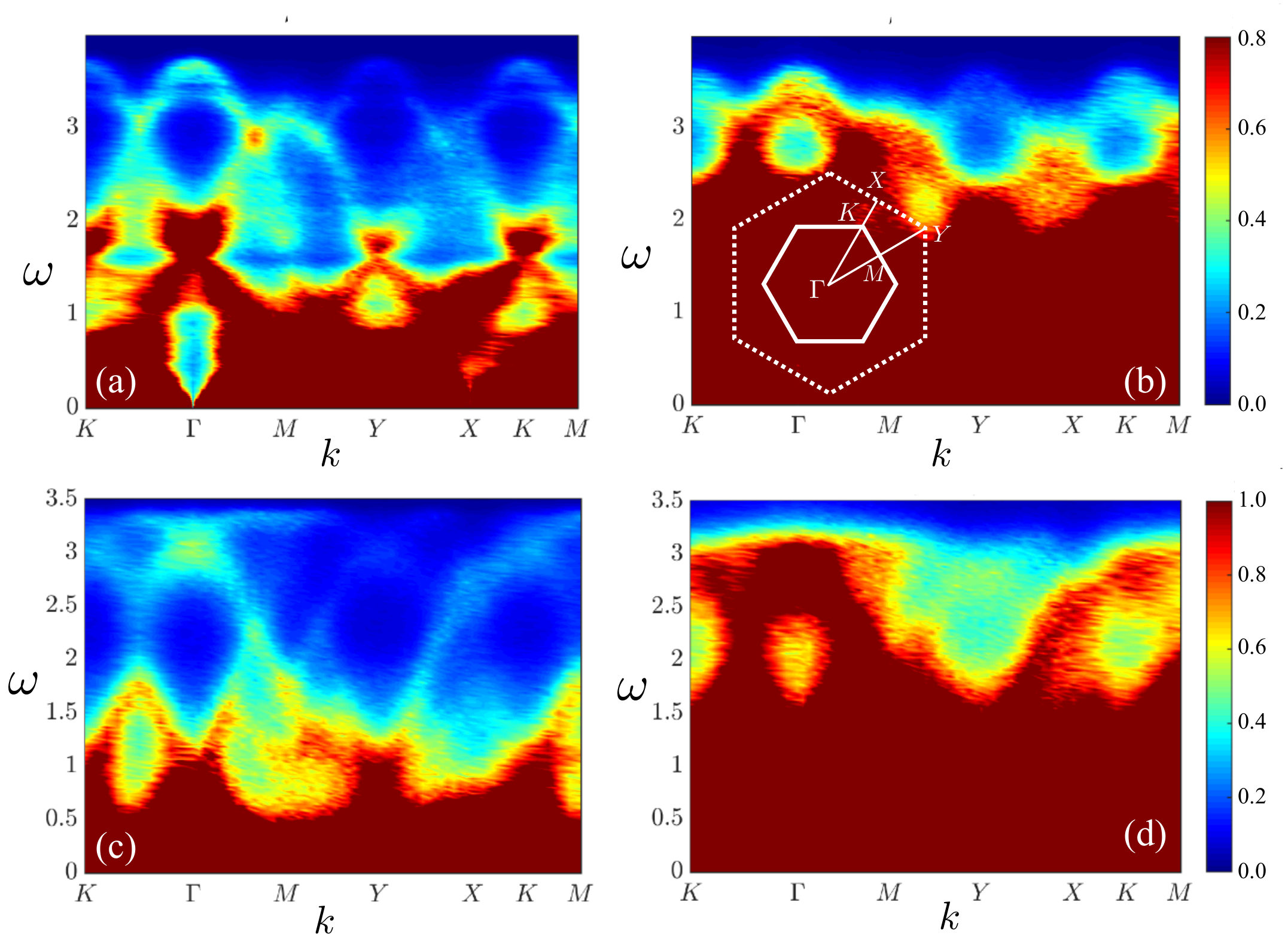}
\caption{(Color online)  
\label{fig:dyn_sf} Dynamical structure factor $S(\mathbf q, \omega)$ computed from LL simulations for antiferromagnetic (top) and ferromagnetic (bottom) $\Gamma$ model. The results below $T_c$ are shown in panels (a) and (c), while those above $T_c$ are shown in (b) and (d).
}
\end{figure}

We next investigate the dynamical behaviors of the spin liquids above and below the critical $T_c$. To this end, we employ the semiclassical Landau-Lifshitz (LL) dynamics simulation, which has been successfully applied to compute the dynamical structure factor $S(\mathbf q, \omega)$ of various classical spin liquids~\cite{moessner98,taillefumier14,samarakoon17}. For $T > T_c$, MC simulations are used to prepare initial states sampled from the Boltzmann distribution. We then perform energy-conserving LL simulation to obtain trajectories of spins $\mathbf S_i(t)$. The dynamical structure factor is computed from the Fourier transform of the real-space correlator $\langle \mathbf S_i(t) \cdot \mathbf S_j(0) \rangle$ averaged over the initial states. As discussed above, since ground states parameterized by different $\{\eta_\alpha\}$ are disconnected below $T_c$, an additional average over random~$\{\eta_\alpha\}$ is introduced manually to improve the efficiency. It is worth noting that the dependence of $S(\mathbf q, \omega)$ on temperature mainly comes from different initial state sampling.

The dynamical structure factor in the two spin liquid phases are shown in Fig.~\ref{fig:dyn_sf} for both signs of $\Gamma$. The $S(\mathbf q, \omega)$ at $T > T_c$ shows broad continuum over a wide energy range in both cases. On the other hand, structures of coherent quasi-particle dispersion can be seen at high energies for $S(\mathbf q, \omega)$ in the low-$T$ phase. These coherent excitations in a liquid phase are reminiscent of the electron pseudo-bands observed in liquid metals~\cite{baumberger04,kim11}. Their origin can be traced to the robust local ordering in a liquid state. Interestingly, the overall results are similar to those observed in the kagome Heisenberg antiferromagnet~\cite{taillefumier14}, in which thermal order-by-disorder induces a coplanar spin liquid phase at low temperatures~\cite{chalker92,chern13}.

{\em Discussion and Outlook}. We have demonstrated that thermal order-by-disorder in honeycomb $\Gamma$-model drives a phase transition into a new spin liquid phase with a hidden flux long-range order. The same scenario also applies to quantum order-by-disorder which generates a similar effective cubic anisotropy~\cite{rousochatzakis17}. In general, we expect this flux-ordered spin liquid to be a strong candidate for low-temperature phases of magnets with a dominant $\Gamma$ interaction, even in the presence of other perturbations. A possible situation is that the flux-ordering is induced energetically by residual perturbations instead of thermal fluctuations. Experimentally, one manifestation of flux ordering is the onset of spin cubic anisotropy. However, this signal might be difficult to detect given the intrinsic cubic symmetry of the system. Through coupling to other degrees of freedom in crystal, e.g. spin-lattice coupling, the translation-symmetry breaking could produce Bragg peaks in neutron or X-ray scattering. 

The effects of quantum fluctuations have been extensively discussed in Ref.~\cite{rousochatzakis17}. The relevant energy scale of quantum order-by-disorder is $T^*\sim \mathcal{O}(|\Gamma| S)$ for both the discrete $\eta$ and continuous $\hat{\mathbf n}$ variables~\cite{rousochatzakis17}. As noted in the same study, the quantum-fluctuation induced effective interaction between $\eta$'s remains frustrated for antiferromagnetic $\Gamma$, so a similar flux-ordered spin liquid can be stabilized by pure quantum fluctuations in this case. It is, however, unclear what is the scenario in the ferromagnetic $\Gamma$ model due to the closeness of the two energy scales. Restoring the spin length, the critical temperature for thermal order-by-disorder is $T_c \sim 0.04 |\Gamma| S^2$. Our finding thus ensures the existence of the exotic flux-ordered spin liquid for large $S$ at the temperature window $T^* \lesssim T \lesssim T_c$. Finally, it is also of great interest to study similar flux-ordering in three-dimensional hyper- or stripy-honeycomb lattices where some of the flux variables are defined on extended strings.



\bigskip

{\em Acknowledgements}. The authors thank C. D. Batista, N. Perkins, and I. Rousochatzakis for useful discussions. G.-W.~C. acknowledges support from the Center for Materials Theory as a part of the Computational Materials Science  (CMS) program, funded by the  DOE Office of Science, Basic Energy Sciences, Materials Sciences and Engineering Division. 

\bigskip

{\em Note added}: Upon finishing our manuscript, we became aware that dynamical structure factor of the $\Gamma$-model has also been studied in~\cite{samarakoon18}.

\end{document}